\documentstyle[aaspp4,11pt,psfig,epsf]{article}

\newcommand{\mgh}[1]{ } 
\newcommand{\mhs}[1]{ }

\newcommand{\Msol}{\mbox{M$_\odot$}}
\newcommand{\op}{Ly$\alpha$\ }

\newcommand{\kms}{\, {\rm km \, s}^{-1} }
\newcommand{\kpc}{\, {\rm kpc} }
\newcommand{\cm}{{\, \rm cm} }
\newcommand{\sr}{{\, \rm sr} }
\newcommand{\Hz}{{\, \rm Hz} }
\newcommand{\erg}{{\, \rm erg} }
\newcommand{\Mpc}{\, {\rm Mpc} }

\newcommand{\hi}{\mbox{H{\scriptsize I}}}

\newcommand{\civ}{\mbox{C{\scriptsize IV}}}

\newcommand{\siii}{\mbox{Si{\scriptsize II}}}

\newcommand{\alii}{\mbox{Al{\scriptsize II}}}
\newcommand{\niii}{\mbox{Ni{\scriptsize II}}}
\newcommand{\feii}{\mbox{Fe{\scriptsize II}}}

\newcommand{\dd}{\,{\rm d}}

\newcommand{\plotmgh}[8]{\vspace{#4}
\hspace{#5}\psfig{file=#1,width=#2,angle=#3}
\vspace{#6}
\caption{\small{#8}}}  
\newcommand{\mclearpage}{\clearpage}

\begin{document}

\title{Damped \op absorbers at high redshift --- \\
large disks or galactic building blocks?}

\lefthead{Haehnelt, Steinmetz, \& Rauch}
\righthead{The Origin of Damped Lyman Alpha Absorbers at High Redshift}

\vskip 1.cm

\author{Martin G. Haehnelt\altaffilmark{1}, Matthias Steinmetz\altaffilmark{1,2}, Michael Rauch\altaffilmark{3,4}
}

\altaffiltext{1}{Max-Planck-Institut f\"ur Astrophysik, Postfach 1523, 85740
Garching, Germany}
\altaffiltext{2}{Steward Observatory, University of Arizona, Tucson, 
AZ 85721, USA}  
\altaffiltext{3}{Astronomy Department, California Institute of Technology,
Pasadena, CA 91125, USA}
\altaffiltext{4}{Hubble Fellow}
\vskip 1.cm

\centerline{mhaehnelt@mpa-garching.mpg.de, msteinmetz@as.arizona.edu, mr@astro.caltech.edu}

\vspace{1.5cm}


\begin{abstract}
We investigate the nature of the physical structures giving rise to
damped \op absorption systems (DLAS) at high redshift. In particular, we
examine the suggestion that rapidly rotating large disks 
are the only viable explanation for the characteristic observed  
asymmetric profiles of  low ionization absorption lines. 
Using hydrodynamic simulations of galaxy formation in a 
cosmological context we 
demonstrate  that irregular protogalactic clumps   
can reproduce the observed velocity width distribution and asymmetries 
of the absorption profiles equally well. The velocity broadening in
the simulated clumps is due to a mixture of rotation, random motions, 
infall and merging. The observed 
velocity width correlates with the virial velocity of the dark matter 
halo of the forming protogalactic clump ($\Delta v \approx 0.6\; v_{\rm
vir}$ for the median values, with a large scatter of order 
a factor two between  different lines-of-sight). 
The typical virial velocity of the  halos required to give rise  to
the DLAS population  is about  $100 \kms$ and most standard 
hierarchical structure formation scenarios can 
easily account even for the largest observed
velocity widths.  We conclude that the  evidence  that DLAS 
at high redshift are related to large rapidly  rotating disks 
with $v_{\rm circ} \ga 200 \kms$ is not compelling.

\end{abstract}

\keywords{galaxies: kinematics and dynamics ---
galaxies: structure --- intergalactic medium --- quasars: absorption lines}

\pagebreak

\section{Introduction}

Damped \op absorption systems (DLAS) have often  been interpreted as
large high-redshift progenitors of present-day spirals which have
evolved little apart from forming  stars (Wolfe 1988; Lanzetta et
al.~1991; Wolfe 1995; Wolfe et al.~1995; Lanzetta, Wolfe \& Turnshek
1995).  A number of observational results have been quoted as being in
support of this hypothesis (see section 6 below). Most recently,
Prochaska \& Wolfe (1997) have investigated a variety of idealised
models  for the spatial distribution and kinematics of the absorbing
gas to test whether they could produce  the  absorption line profiles
of low ionisation ionic species (LIS) associated with DLAS.  Of those
models they investigated,  only the one in which the lines-of-sight
(LOS) intersect  rapidly rotating thick galactic disks can explain both
the large velocity spreads (up to  200kms$^{-1}$) and the
characteristic  asymmetries of the observed LIS absorption profiles.
In particular, they find that if the embed their disk model within a CDM 
structure formation scenario, the result is inconsistent with the  observed
velocity widths. In this paper we demonstrate that the inconsistency
with  galaxy formation models within hierarchical cosmogonies 
(e.g. Kauffmann 1996) disappears if the  gas is modeled with a 
more realistic spatial distribution and kinematic structure.

For this purpose we use numerical simulations of galaxy formation 
in a CDM cosmogony including gas dynamics and realistic initial
conditions. These exhibit  a complex relationship between high 
column density absorption features and the underlying dark matter 
distribution (Katz et al.~1996; Haehnelt, Steinmetz \& Rauch  1996, 
paper I; Rauch, Haehnelt \& Steinmetz 1997, paper II; Gardner et
al. 1997a/b).  Agglomerations of
neutral hydrogen with central column densities larger than
$10^{20}\cm^{-2}$ and with the masses of dwarf galaxies do 
occur commonly in these simulations. These objects form by gravitational collapse
in CDM potential wells.  Subsequent cooling produces an optically
thick, mostly neutral phase in the inner ten to twenty  kpc.  
We have already demonstrated that the large number
of these objects and their clustering and merging into larger units can
explain many observed features of metal absorption systems, for
example the ionization and thermal state of the gas, and the observed
multi-component structure of the absorption line profiles (paper I\&II).  
In these models the high rate of incidence of damped \op systems 
is a result of the high abundance of protogalactic clumps (PGC) 
which are the progenitors of large present-day  galaxies.
This must be contrasted with the popular picture of DLAS where a population
of very large disks evolves  without merging to form present-day 
spirals.

Prochaska \& Wolfe (1997) have highlighted two crucial questions which
a hierarchical structure formation model must be able to address
satisfactorily: 

\begin{itemize}
\item[(1)] How do the observed asymmetries of the absorption line
complexes  arise?
\item[(2)] Can  absorption by groups of PGCs in a hierarchical universe 
reproduce the observed velocity width distribution? 
\end{itemize}

Below we will investigate the velocity width and shape of LIS absorption 
profiles  using  artificial spectra  for lines-of-sight through
numerically simulated regions of ongoing galaxy formation. 
We then examine the underlying
physical conditions responsible for the kinematic structure of these
systems. We further  investigate the connection between the velocity 
width of the absorption systems and the depth of the forming potential
well and assess the problem of accommodating the observed velocity 
width within standard hierarchical cosmogonies. Finally we  discuss
our results and some other, observational clues to the nature
of DLAS, and draw conclusions.

\section{Numerical simulations of damped \op systems}

\subsection{The hydrodynamical simulations}

Spatial regions of the universe selected to contain  one or a few
normal galaxies at redshift zero are simulated with the
hydrodynamic GRAPE-SPH code (Steinmetz 1996) in the framework of a 
standard CDM cosmogony ($\sigma_{8} = 0.67$, $H_{0} = 50\kms\Mpc^{-1}$,
$\Omega_{\rm b} = 0.05$). 
Temperature, density and peculiar velocity arrays along LOS through the
simulated boxes are used to produce artificial absorption spectra.  
For a detailed description of the properties of the simulations and the
resulting absorption features see Steinmetz (1996), 
Navarro \& Steinmetz (1997), and papers I\&II.  The strategy of simulating  
small regions of ongoing galaxy formation preselected from a large 
dark matter simulation allows us to achieve a spatial resolution of 1\,kpc and 
a mass resolution of $5\times 10^6\,\Msol$ (in gas). This high  resolution 
--- about a factor ten higher than that in most other cosmological 
hydro-simulations  --- is crucial to resolve the rich
substructure within the damped region induced by the frequent merging 
of protogalactic clumps in hierarchical  structure formation
scenarios. Despite this, the resolution is still not  sufficient to 
account for a possible clumping on sub-kpc scales due to thermal instabilities 
(Mo 1994; Mo \& Miralda-Escud\'e 1996).  Furthermore energy and 
momentum feedback  due to star formation are not  included. Both
effects are likely to produce additional substructure in physical as
well as in velocity space.

\subsection{The region of neutral hydrogen --- self-shielding}

To study the kinematic structure of damped systems we extend  our
previous work  to LOS passing through  regions of collapsed 
dark matter halos with integrated \hi \ column densities 
exceeding $2 \times 10^{20}$ cm$^{-2}$. The main problem which then arises  
is  the treatment of the self-shielding of the dense gas against  radiation beyond 
the Lyman edge. With the current generation of computers it is not yet 
possible to run cosmological hydro-simulations which solve the full 
radiative transfer equations. We therefore have adopted a simple scheme to 
mimic the effect of self-shielding which is motivated by the 
tight correlation between  column density and density 
predicted by the numerical simulations (Miralda-Escude et al 1996; 
paper II).  A \hi\ column density of $10^{17}\cm^{-2}$, above 
which self-shielding  becomes important, occurs for LOS with an 
absorption-weighted density of about $10^{-3} \cm
^{-3}$ to $10^{-2} \cm ^{-3}$. This is easy  to understand by 
looking at  the photoionization equilibrium equation for a 
highly ionized optically thin homogeneous slab of hydrogen. 
The column density  of neutral hydrogen then scales 
as $N_{\rm HI} \propto n_{\rm H}^{2}\, D\,J_{912}^{-1}$,     
where $D$ is the thickness of the slab and $J_{912}$  is the flux of 
the UV background at the Lyman edge.            
The hydrogen density at the onset of self-shielding in the central plane 
can  be written as, 
\begin{equation}
n_{\rm shield} \sim
        3 \times 10^{-3} \;
        \left ( \frac{D}{10 \kpc} \right )^{-0.5} \;
        \left (\frac{J_{912}}{0.3 \times 10^{21}
        \erg\cm^{-2}\sec^{-1}\Hz^{-1}\sr^{-1}}
        \right )^{0.5}
        \cm^{-3},
\end{equation}
where the spectral shape proposed by Haardt \& Madau (1996) was used to
transform the  flux at the Lyman edge $J_{912}$  into a
photoionization rate. The geometry of  collapsed regions in the  
numerical simulations is certainly more complicated than that of a 
slab, but $D$ = 10 kpc is a typical scale. To be on the safe side we 
assumed  that all the gas above a density threshold of 
$10^{-2}\cm ^{-2}$ is self-shielded. This probably underestimates the 
size of the self-shielding region.

\subsection{Profiles of low ionization ionic species}

In observed DLAS hydrogen is predominantly neutral 
due to the self-shielding of the gas while the other atomic
species attain low ionization states (Viegas 1995).  
Optically thin transitions of LIS like  \siii\, \alii\ , \feii\ and 
\niii\  are therefore generally considered as  suitable  
tracers of the  the kinematic and 
density structure of the neutral gas (Wolfe 1995). We
have chosen  the \siii\ 1808 absorption feature for our investigation. 
Silicon was assumed to be  predominantly in the first ionization 
state within the self-shielding region, 
\begin{eqnarray*}
\nonumber \frac{[\siii]}{\rm [Si]} = 1, &n_{\rm H} >10^{-2} \cm^{-3}, & \\
\nonumber &&\\
\nonumber \frac{[\siii]}{\rm [Si]} = 0, &{\rm otherwise}.& \\ 
\end{eqnarray*}
A homogeneous  silicon abundance of  [Si/H] = $-1$ 
was assumed for the self-shielding region.  
Artificial spectra were
made to resemble typical Keck data obtainable within a few hours from a
16-17th  magnitude QSO (S/N= 50 per 0.043 \AA\ pixel, FWHM
= 8km/s). 
To make contact with our previous work we will also show 
the corresponding \civ\ absorption line profiles. For these 
a homogeneous  carbon abundance of  [C/H] = $-1.5$ was assumed.
The \civ\ fraction was calculated using the photoionization code 
CLOUDY (Ferland 1993) as described  in papers I\&II. 

\subsection{A gallery of simulated damped \op absorbers}

Figs 1-5 present  some typical examples of simulated damped \op absorption
systems. The {\it bottom left} panel shows the absorption spectrum for
the \siii\ 1808  and  \civ\ 1548 
transitions, the {\it top left} and {\it bottom right} show 
the total hydrogen density  and peculiar velocity along the LOS, while
the {\it top right} panel shows density and velocity fields in a thin 
slice containing the LOS.  The coordinates are proper
distance, and the projection is such that the LOS 
along which the absorption spectrum is determined  lies 
along the z-axis. The wavelength axis of the spectra is in $\kms$. 
Velocities are relative to the center-of-mass  velocity of a sphere
with 30 kpc radius. The assumed density threshold 
for self-shielding is indicated by the dashed line in the density
profile and by the thick contour in the slice of the density 
field.

Fig.~1 shows a typical example of two  merging PGCs. The 
density profile along the line of sight has two peaks 
within the self-shielding region. On larger scales the  gas is 
flowing in from the left and the right, but in the self-shielded
region the flow is  rather quiescent with a velocity gradient of 
only about $30 \kms$.  This results in a double-peaked \siii\ 1808 absorption 
profile with about $30\kms$ velocity difference between the peaks.  

Fig.~2 shows another example of  merging PGCs. This time the 
density profile along the line of sight has three 
rather marginal peaks within the self-shielding region. 
The velocity profile is smooth
but shows large gradients due to an eddy-like  motion in the shock  
produced by the infalling clump. The corresponding \siii\ 1808
absorption profile is complex and extends over $200 \kms$, showing a 
prominent leading edge. The hump in the velocity profile at 5kpc and the 
three density peaks show all up as individual absorption features.

\begin{figure}[ht]
\plotmgh{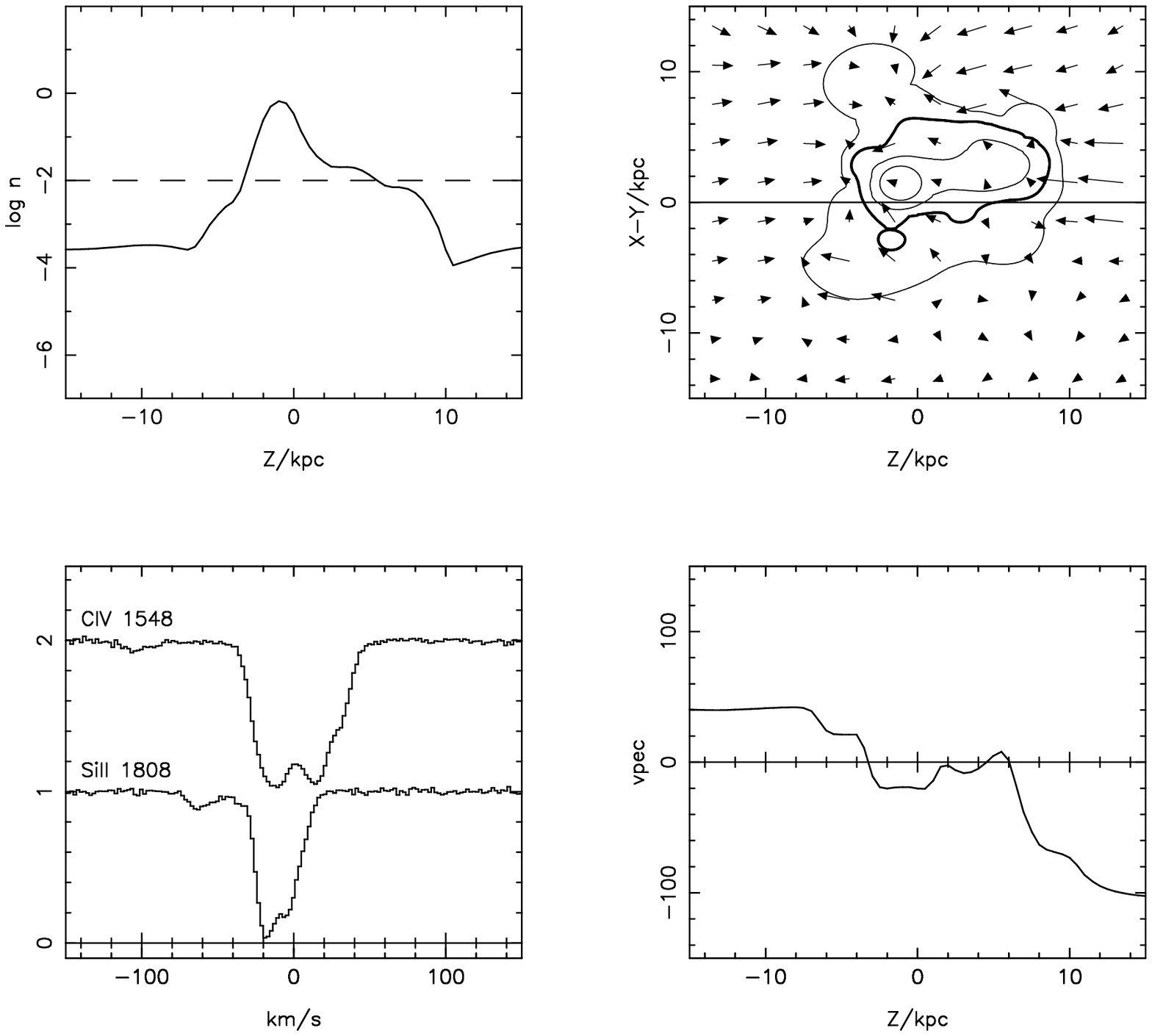}{16.0cm}{0.0}{0.cm}{0.cm}{0.5cm}{1a}
{Simulated damped \op absorber at $z$ = 2.1
         with $\log N_{\hi} = 21.2$ arising from
         gas in a merging protogalactic clump with 
         $v_{\rm vir} = 70 \kms$.
          {\it Bottom left}: Absorption spectrum for the \siii\ 1808
                 and  \civ\ 1548. 
                      transitions. 
         {\it Top left}: Total hydrogen density along the LOS. 
         {\it Bottom right}: Peculiar velocity along the LOS. 
         {\it Top right}: Density and velocity field in a thin 
                    slice containing the LOS (straight solid line).
                    The density contours have a spacing of 1 dex and 
                    the thick contour marks $\log n$ = $-2$. Velocities
                    are relative to the center-of-mass velocity of a sphere
                    with 30 kpc radius. The normalization of  
                    the velocity arrows is such that the length
                    of the longest arrow equals the spacing betwen 
                    arrows. For absolute velocity values see the
                    bottom right panel.}
\end{figure}

\begin{figure}[ht]
\plotmgh{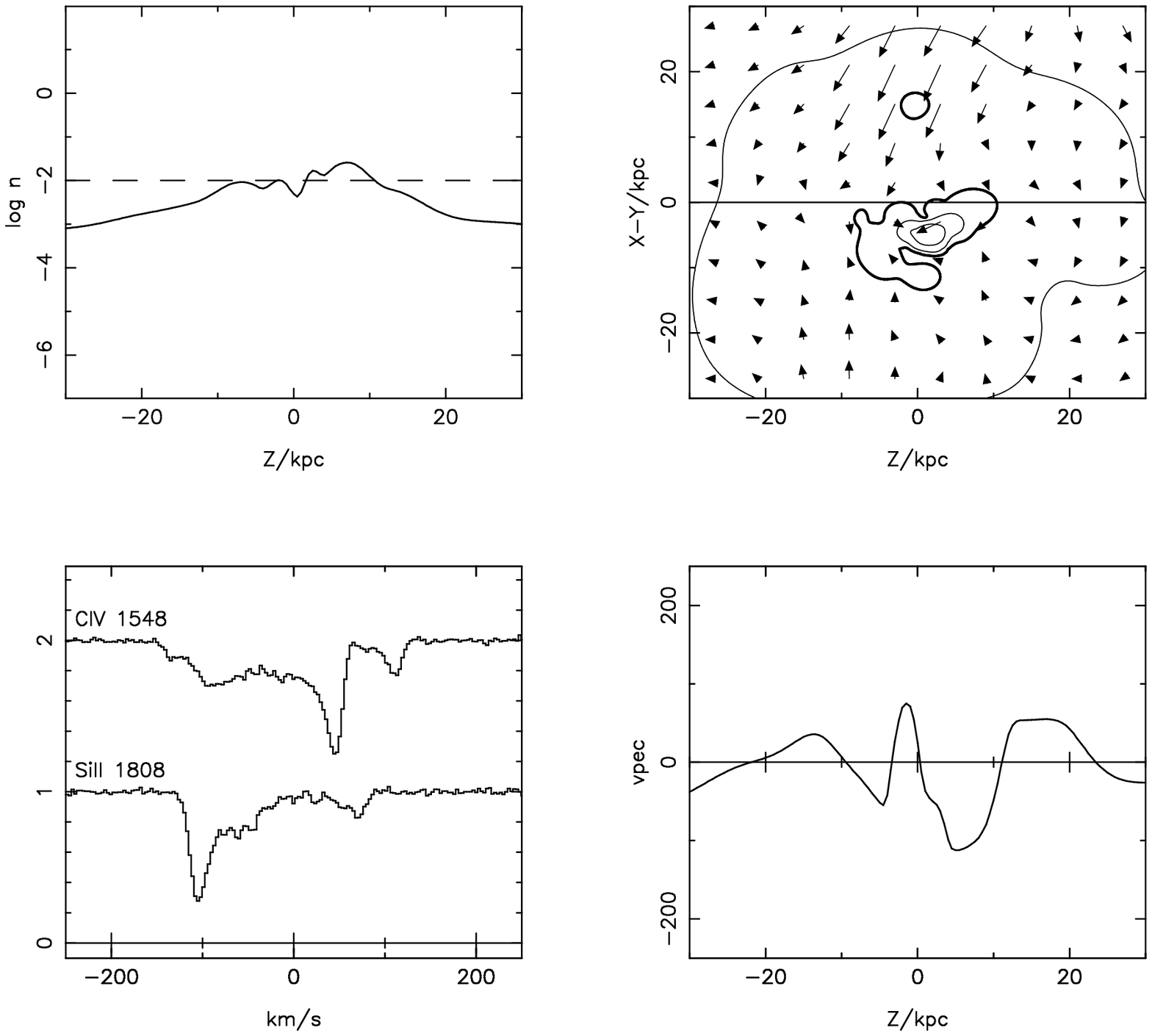}{16.0cm}{0.0}{0.cm}{0.cm}{0.5cm}{1b}
{Simulated damped \op  absorber at $z$ = 3.1 with $\log N_{\hi} = 20.6$
arising from gas in a merging protogalactic clump with 
$v_{\rm vir} = 65 \kms$.
}
\end{figure}

\begin{figure}[ht]
\plotmgh{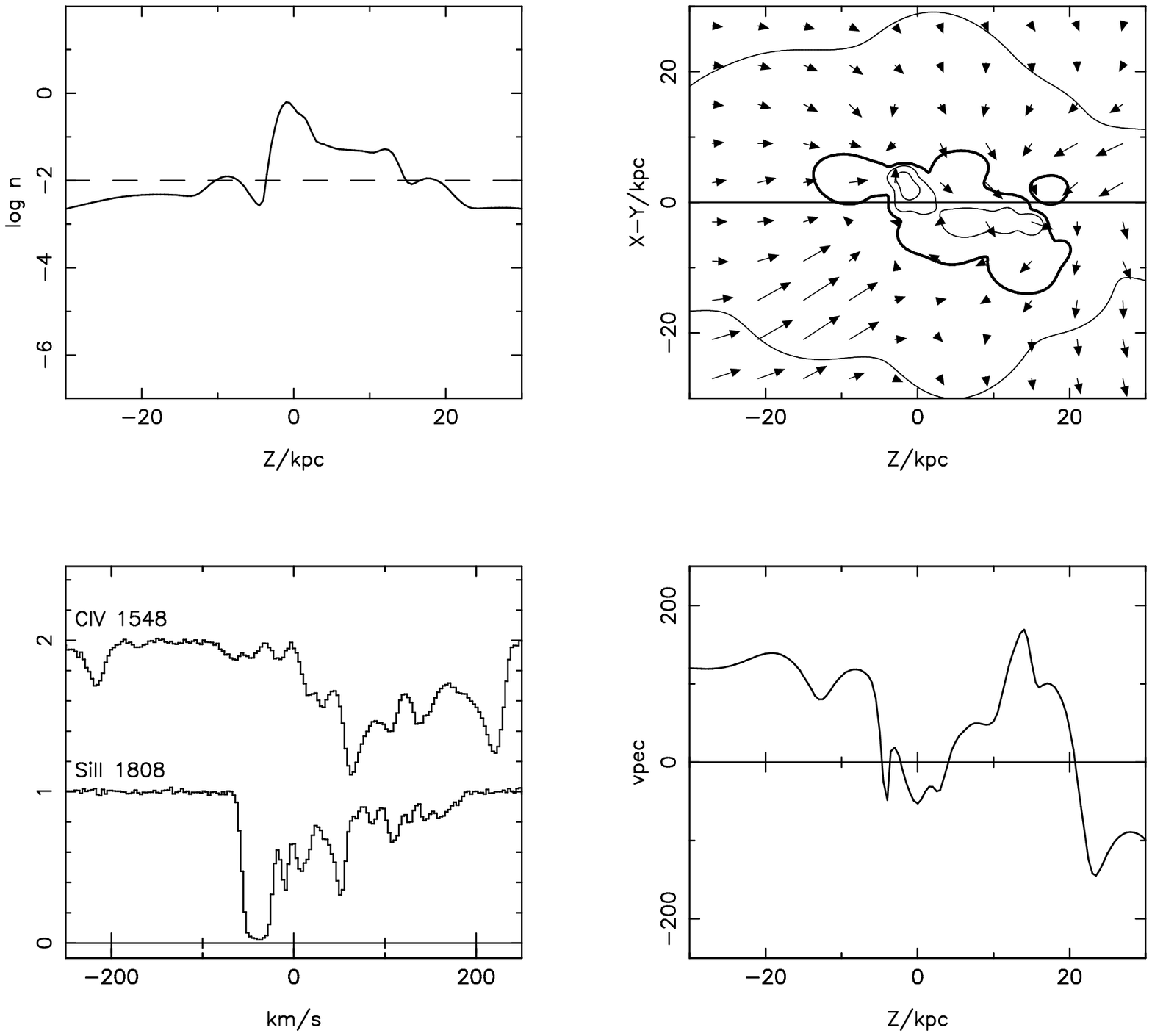}{16.0cm}{0.0}{0.cm}{0.cm}{0.5cm}{1c}
{Simulated  damped \op absorber  at $z$ = 3.1
with $\log N_{\hi}$ = 21.7 arising from
gas in a merging protogalactic clump 
with $v_{\rm vir}$ = $200 \kms $.
}
\end{figure}

\begin{figure}[ht]
\plotmgh{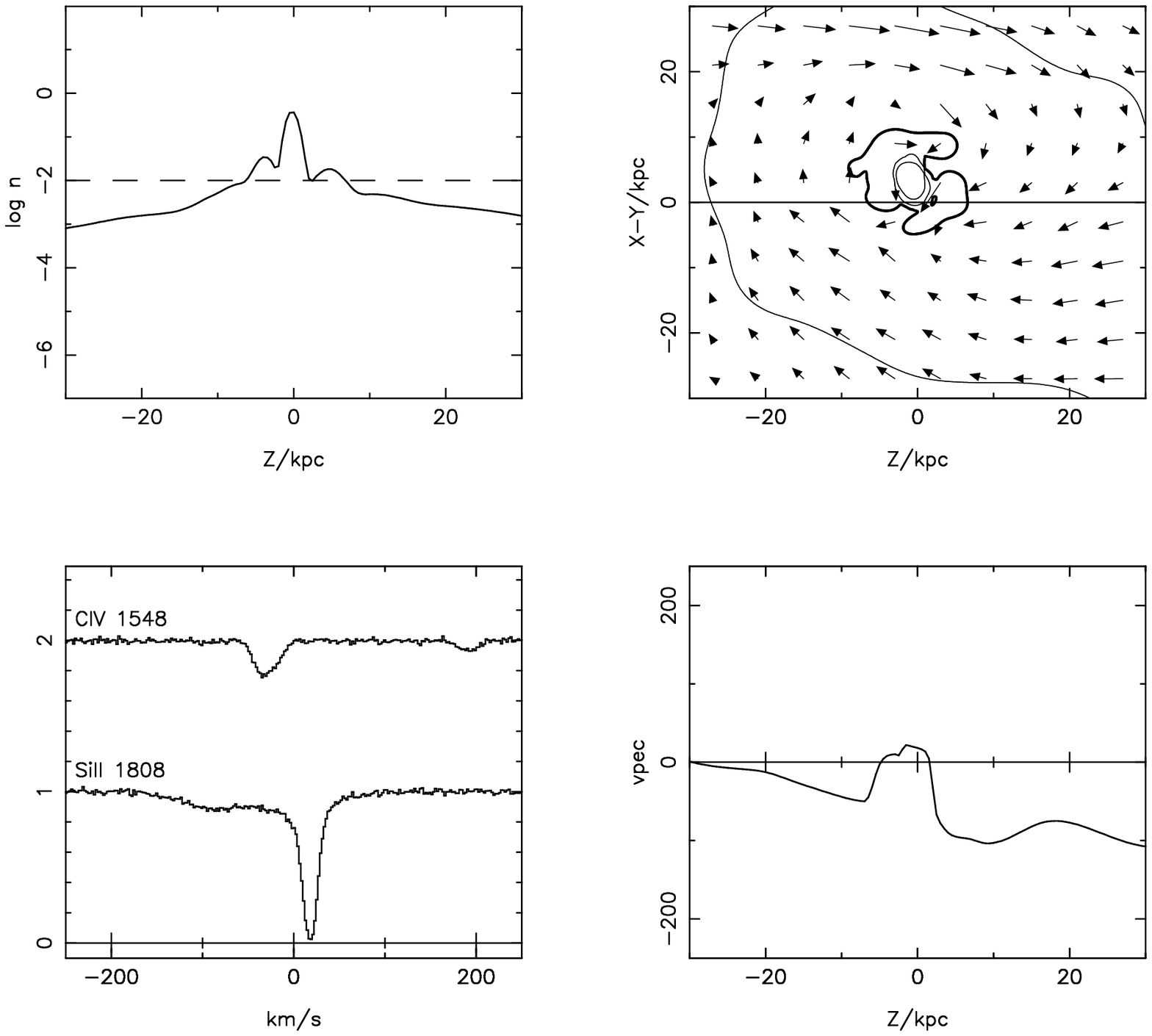}{16.0cm}{0.0}{0.cm}{0.cm}{0.5cm}{1d}
{Simulated damped \op  absorber at $z$ = 2.1 with $\log N_{\hi} = 21.3$ 
         arising from gas in a ``rotating'' protogalactic clump with 
         $v_{\rm vir} = 180 \kms $.}
\end{figure}

\begin{figure}[ht]
\plotmgh{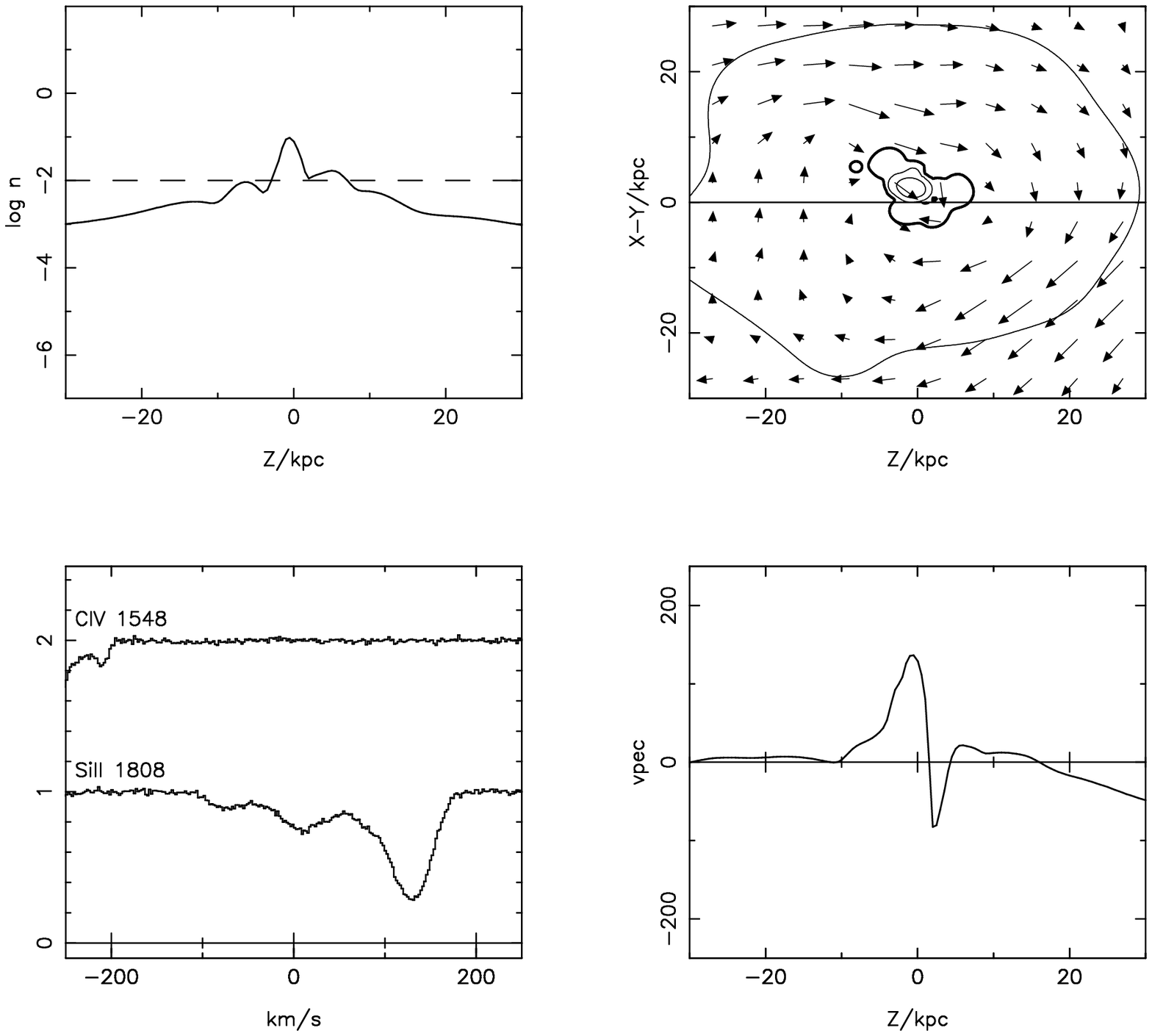}{16.0cm}{0.0}{0.cm}{0.cm}{0.5cm}{1e}
{Simulated damped \op  absorber at $z$ = 2.1 with $\log N_{\hi} = 20.8$ 
         arising from gas in a ``rotating'' protogalactic clump with 
         $v_{\rm vir} = 180 \kms $.}
\end{figure}

\vfill
\mclearpage

Fig.~3 shows a similar third example of merging PGCs. The peculiar velocity
profile shows the same large gradients but is much more chaotic than 
that in Fig.~2. The peaks of the density profiles and the features of
the velocity profile in the self-shielding region can again be
identified  in the \siii 1808 absorption profile which also shows 
a prominent  leading edge. One should note here the strong difference
between the \siii\ 1808 profile and the  \civ\ 1548 profile.
The latter arises from absorption by 
the spatially separated warm gas surrounding the self-shielding region.

Fig.~4 and 5 show two of the rather rare cases where there is a large
rotational component in  the motion of the gas. For these ``rotating''
PGCs  we  generally find rather smooth density and velocity profiles in the 
self-shielding region.  In most cases these result in  single-peaked 
\siii\ 1808 absorption  profiles with mild asymmetries or extended
wings to one side as in Fig.~4. The latter are often difficult to
detect unless the main peak is already saturated.  The best example 
of a leading edge profile produced by rotation which we could find
is shown in Fig.~5.


\subsection{Orientation effects}

As discussed above,  the detailed structure of the
absorption profile of the LIS regions depends strongly on the
substructure in physical and velocity space. This makes the 
absorption profiles very sensitive to the orientation of the 
LOS. In order to demonstrate this  we plot in  Fig.~6 ten different randomly
oriented LOS, each giving rise to damped \op absorption, in the vicinity of 
the PGC shown  in Fig.~1. The nature of the absorption
profiles varies from a single symmetric peak to double and multiple 
peaks. The rapid changes in the details of the velocity profile along 
the  LOS are due to the rather chaotic velocity field of the merging 
PGC, and are  the main reason for the dramatic changes in the LIS absorption 
profile. It is also clearly seen that the absorption profile 
of the higher ionization ion \civ\ varies independently of \siii . 
This is due to the fact that \civ\ arises mainly from the warmer gas
outside the self-shielding region.

\begin{figure}[ht]
\plotmgh{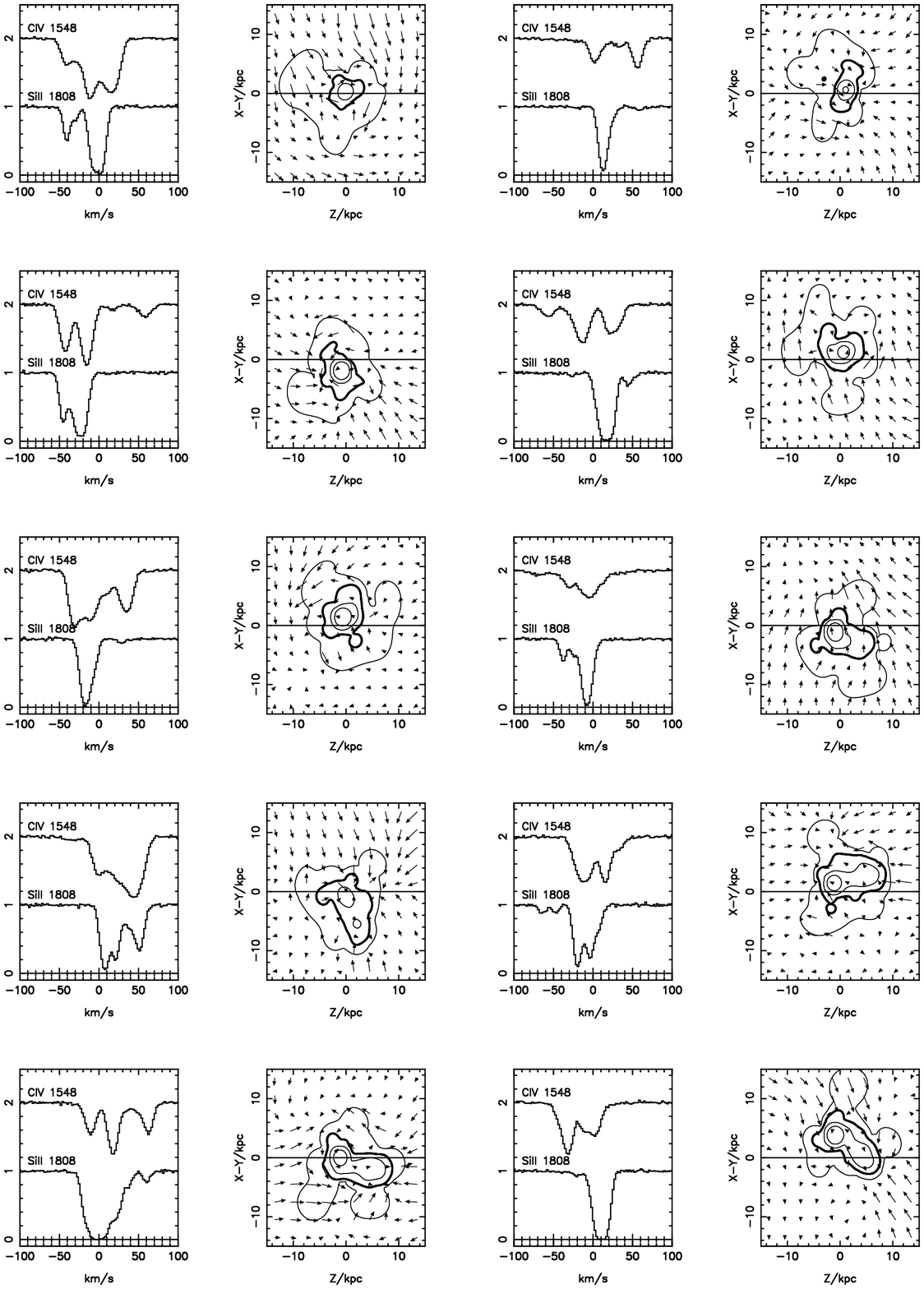}{15.0cm}{0.0}{0.5cm}{0.5cm}{0.5cm}{2}
{Ten random LOS producing damped \op in the vicinity 
         of the merging protogalactic clump shown in Fig.~1. Plotted are 
         absorption spectra and corresponding density/velocity fields
          as in the top right and bottom left panels of Fig.~1.}

\end{figure}

\vfill
\mclearpage

\begin{figure}[ht]
\plotmgh{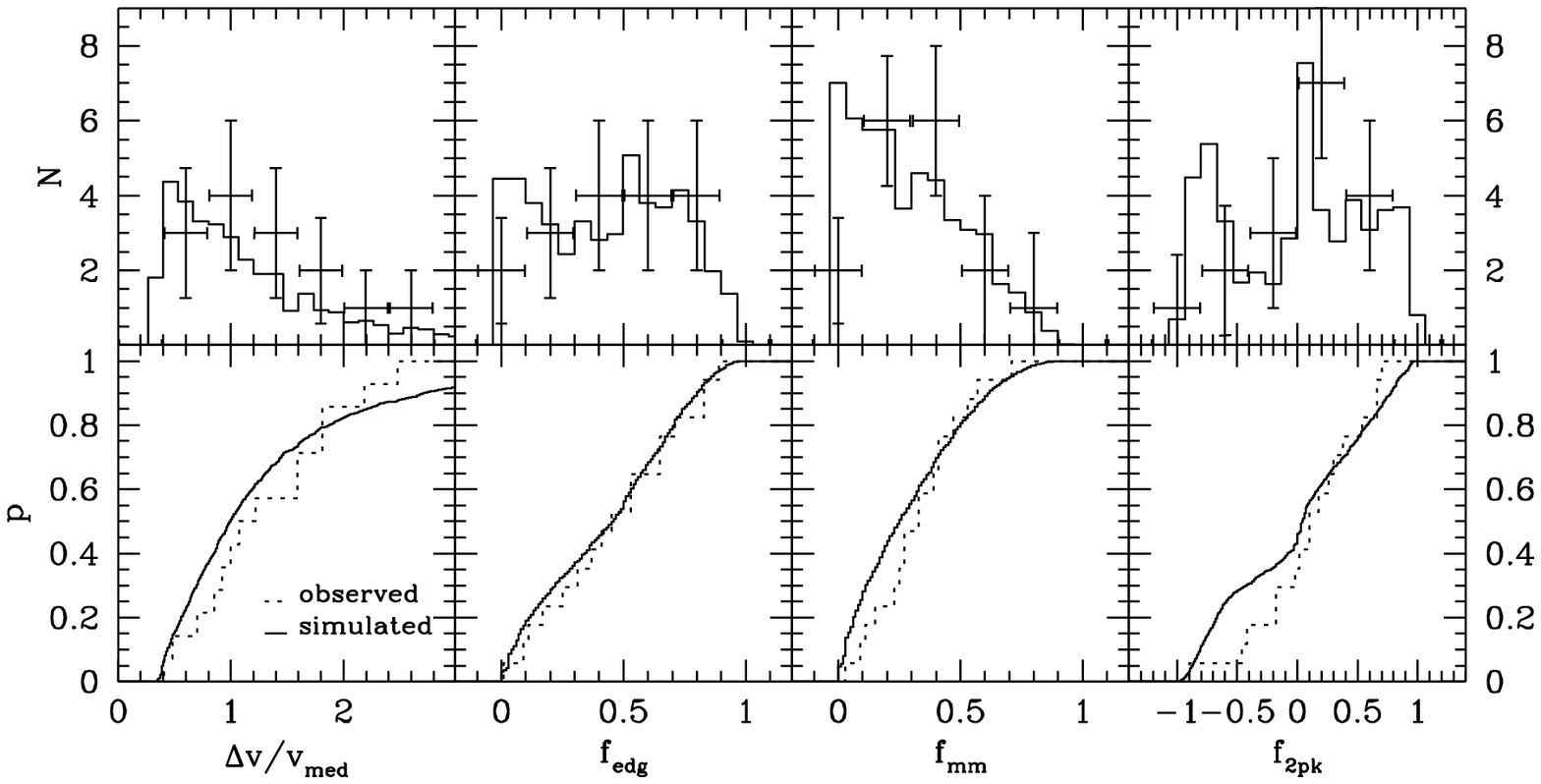}{17.0cm}{0.0}{-1.0cm}{-1.5cm}{-8.5cm}{3}
{Velocity width distribution and shape parameter ---
From left to right: velocity width (relative to the median 
value),  edge-leading index, mean-median index and
2-peak index. The top panels show the differential distribution
in absolute numbers (model curves normalized to the observed number of
systems); the bottom panels show the cumulative probability distribution.
The data is taken from Prochaska \& Wolfe (1997).}

\end{figure}

\section{Distribution of velocity widths and shapes --- Observed 
            {\it vs} simulated}

Prochaska \& Wolfe (1997) have introduced four parameter to characterize 
LIS absorption profiles. In order to facilitate comparison with 
their observed sample of DLAS  (a sample of 17 DLAS obtained by Sargent
et al. with the Keck HIRES instrument)
we have  applied their selection criteria and 
characterized the simulated LIS profiles in exactly the 
same manner.  The \siii\ 1808 absorption profiles were
transformed  into an apparent optical depth profile which was then 
smoothed over a range of 9 pixels.  
The first parameter ---  the velocity width $\Delta v$ of the LIS
region --- is defined as the  velocity interval which contributes
the central 90 percent to the optical depth weighted velocity integral, 
\begin{equation}
\tau_{\rm tot} = \int{ \tau \dd v.}
\end{equation}
The mean velocity $v_{\rm mean}$ is defined as the midpoint of the
velocity width interval (setting $v$ = 0 at the left edge) 
while the median velocity $v_{\rm med}$ bisects the integral in 
equation (2) performed over the velocity width interval. 
The three shape  parameters designed to detect asymmetries in the  
absorption complexes are defined as follows,
\begin{equation}
f_{\rm edg} = \frac{|v_{\rm pk} - v_{\rm mean} |}{(\Delta v/2)},
\end{equation}
\begin{equation}
f_{\rm mm} = \frac{|v_{\rm median} - v_{\rm mean} |}{(\Delta v/2)},
\end{equation}
\begin{equation}
f_{\rm 2pk} = \pm \frac{|v_{\rm 2pk} - v_{\rm mean} |}{(\Delta v/2)},
\end{equation}
where  $v_{\rm pk}$ and $v_{\rm 2pk}$ are the velocity 
of the highest and second highest significant peak in the smoothed 
apparent optical depth profile.  For the 2-peak test the plus (minus) sign
holds if the velocity of the second peak falls (falls not) 
between the velocity of the first peak and the mean velocity. 
In the case of single peaks  $f_{\rm 2pk}$ is set equal to $f_{\rm
edg}$.  To avoid saturation 
effects and to ensure sufficient signal-to-noise only absorption
profiles  with 
\begin{equation}
0.1 \le \frac{I_{\rm pk}}{I_{0}}\le 0.6
\end{equation}
were considered, where $I_{\rm pk}$ and $I_{0}$ are the intensity 
of the strongest peak and the continuum respectively. 

A sample of 640 simulated damped \op absorption systems 
with column densities  above $2\times 10^{20} \cm^{-2}$ and satisfying the
criterion in equation (6) was assembled by choosing random LOS 
in the vicinity of 40 protogalactic clumps  identified with a 
friends-of-friends  group-finder. 
For the discussion of velocity widths a minimum  threshold of 
$\Delta v$ $>$ 30 kms$^{-1}$  was imposed on both the
observed and simulated velocity widths
to avoid incompleteness effects.  The redshift of the simulated
DLAS is $z$ = 2.1, the median redshift of the observed sample.

\begin{deluxetable}{cccccc}
\small
\tablewidth{0pt}
\tablenum{1}
\tablecaption{KS tests observed vs simulated.}
\tablehead{
\colhead{} &
\multicolumn{4}{c}{KS Probabilities for P\&W Tests} &
\colhead{} 
\\
\tablevspace{.2cm}
\cline{1-5}\\
\tablevspace{-.3cm}
\colhead{}&  
\colhead{$P_{\Delta v}$} &
\colhead{$P_{\rm edg}$} &
\colhead{$P_{\rm mm}$} &
\colhead{$P_{\rm 2pk}$} &
\colhead{}  \\
\tablevspace{-.4cm}
}
\startdata
&0.69 & 0.98& 0.23 & 0.28& \\
\enddata
\end{deluxetable} 

Fig.~7 summarizes the comparison between the observed and simulated 
samples. The top panels show the differential distributions 
of the velocity width, the edge-leading, the mean-median,
and the 2-peak parameter. The observed sample is still  small and 
we therefore  had to use rather large bins for the observed data.
The absolute numbers of the observed sample are plotted  
while  the simulated sample is normalized accordingly. The error bars
indicate the Poisson errors and the width of the bin respectively.  
The velocity width is plotted relative to the median value  of the
observed sample and relative to the median value of subsamples 
of 16 LOS around each PGC  in the simulated sample. This allows us to test the
relative velocity distribution independent of the
cosmological  model chosen. We come back to  the absolute velocity width
of the LIS region in section 5.  The bottom panels show the
corresponding cumulative distribution. The agreement between 
observed and simulated spectra is within the expected statistical 
errors. The KS test values for the cumulative distribution are given
in Table 1. For none of the parameters is the KS test probability 
smaller than 20\%. We would, however, like to  caution against
using KS tests  to discriminate between different models. 
Small KS probabilities can be very misleading if a very special 
representation of a general class of models is chosen.
We have e.g. varied our  density threshold 
for the self-shielding region, the redshift of the sample and  the
metallicity of the gas and found significant changes in the distribution of
the parameter. In some cases they further improved the agreement, and
in some cases they led to KS probabilities  as small as a tenth of a
percent for one or two  of the parameters. We believe that a 
significantly larger sample and a careful assessment of the selection
effects are necessary in order to draw strong conclusions from the detailed 
distributions of the shape parameters introduced by Prochaska \& Wolfe.

\section{Physical conditions giving rise to damped Lyman $\alpha$ systems}

In the previous section we have shown that LOS passing the vicinity
of  PGCs can give rise to DLAS with LIS absorption profiles  which reproduce 
the characteristic velocity width distribution  and asymmetries of
observed DLAS.  It remains to be seen which underlying physical
conditions are giving rise to these features. In hierarchical
structure formation scenarios the ``progenitor'' of a present-day
galaxy consists of several PGCs often moving along filamentary
structures  to merge into larger objects.  We found the turbulent gas
flows and inhomogeneous density structures related to the merging  
of two or more  clumps to be the main reason for the occurrence 
of multiple LIS absorption systems with large velocity widths. 
Rotational motions of the gas play only a minor role  for these 
absorption profiles, as does  the velocity  broadening due
to the Hubble expansion between aligned clumps which is 
important for higher ionization species like \civ\ (see papers I\&II).
The latter is easily understandable given the small cross
section of the LIS region.

\subsection{Properties of the absorbing protogalactic clumps} 

We have systematically investigated the physical properties 
of our sample of 40 PGCs in order to understand what kind
of motions are reflected in the LIS line profiles
and how their velocity width is related 
to the depth of the potential wells in which they are embedded.
For this purpose we determined the
following  quantities for the absorbers:
total velocity dispersion of the large scale motions of the gas
in the self-shielding region, velocity dispersion due to radial
motions 
of the gas, 
overall rotational velocity of the  gas and virial velocity  
of the  dark matter halo.  The virial velocity is defined as 
$v_{\rm vir} =  \sqrt{GM/r}$ in
a sphere overdense by a factor 200 compared to the mean cosmic 
density. One should note here that the DM halo(s) are not necessarily
virialized during  the merger of two PGCs.

\begin{figure}[t]
\plotmgh{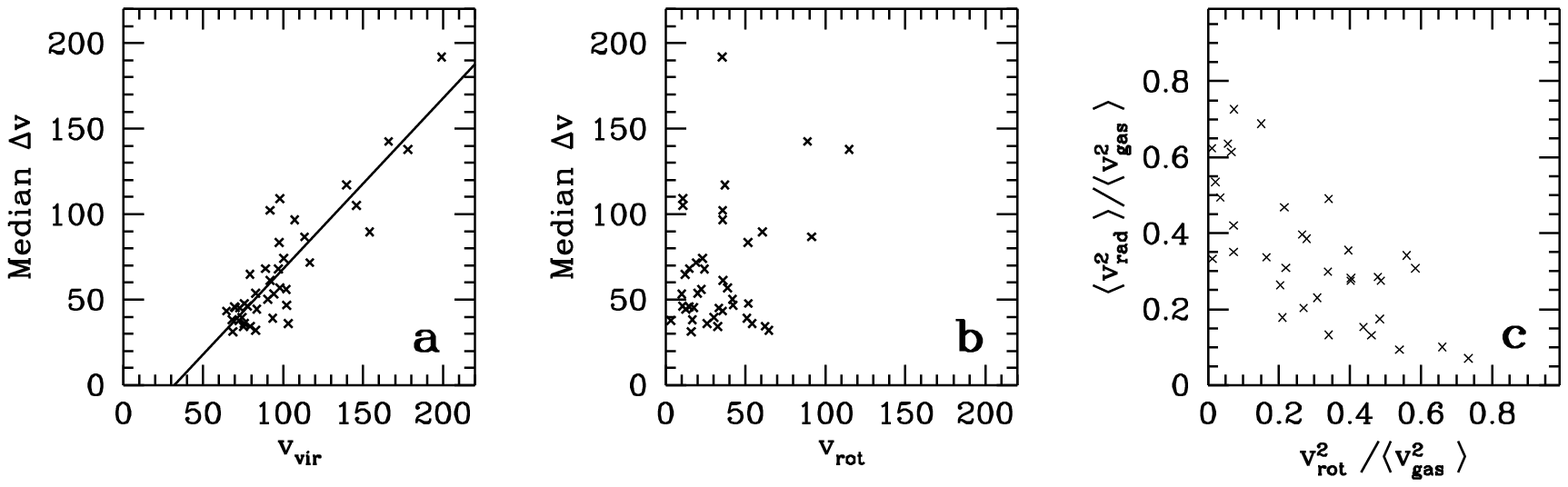}{17.0cm}{0.0}{-2.0cm}{-1.cm}{-12.5cm}{4}
{\hfill\break
a) Median velocity width of the LIS region 
for random LOS around 40 protogalactic clumps (Median of 16 LOS per PGC)  
{\it vs} virial velocity of the associated DM halo. 
\hfill \break    
b) Median velocity width {\it vs} mean rotational velocity  of the gas.
\hfill \break 
c) Relative contributions of rotation and radial infall to the motion
of the gas.\hfill \break}   
\end{figure}

In Fig.~8a  the median value of the velocity width  for 16 randomly 
orientated LOS around each PGC is plotted {\it vs}   
the virial velocity of the DM halo.  There is a strong correlation 
indicating that the velocity width reflects the depth of the 
potential reasonably well even though  there is 
considerable scatter.  The velocity width  of the LIS absorption
region is typically 60 percent of  the virial velocity of the 
DM halo. The solid line shows the least-square fit.

Fig.~8b shows the relation between rotational velocity and  
velocity width. There seems to be no correlation. 
The rotational velocity is generally  too small to account 
for the observed LIS velocity width. In Fig.~8c the relative 
contribution of radial motions and rotation to the total velocity 
dispersion of the gas is shown. 
The contributions of  rotation and radial motions 
(mainly infall and merging) range  between 0 and 70  percent. 
As expected these are anti-correlated. The contribution
of additional random motions is generally between 30 and 70 
percent.

\begin{figure}[t]
\plotmgh{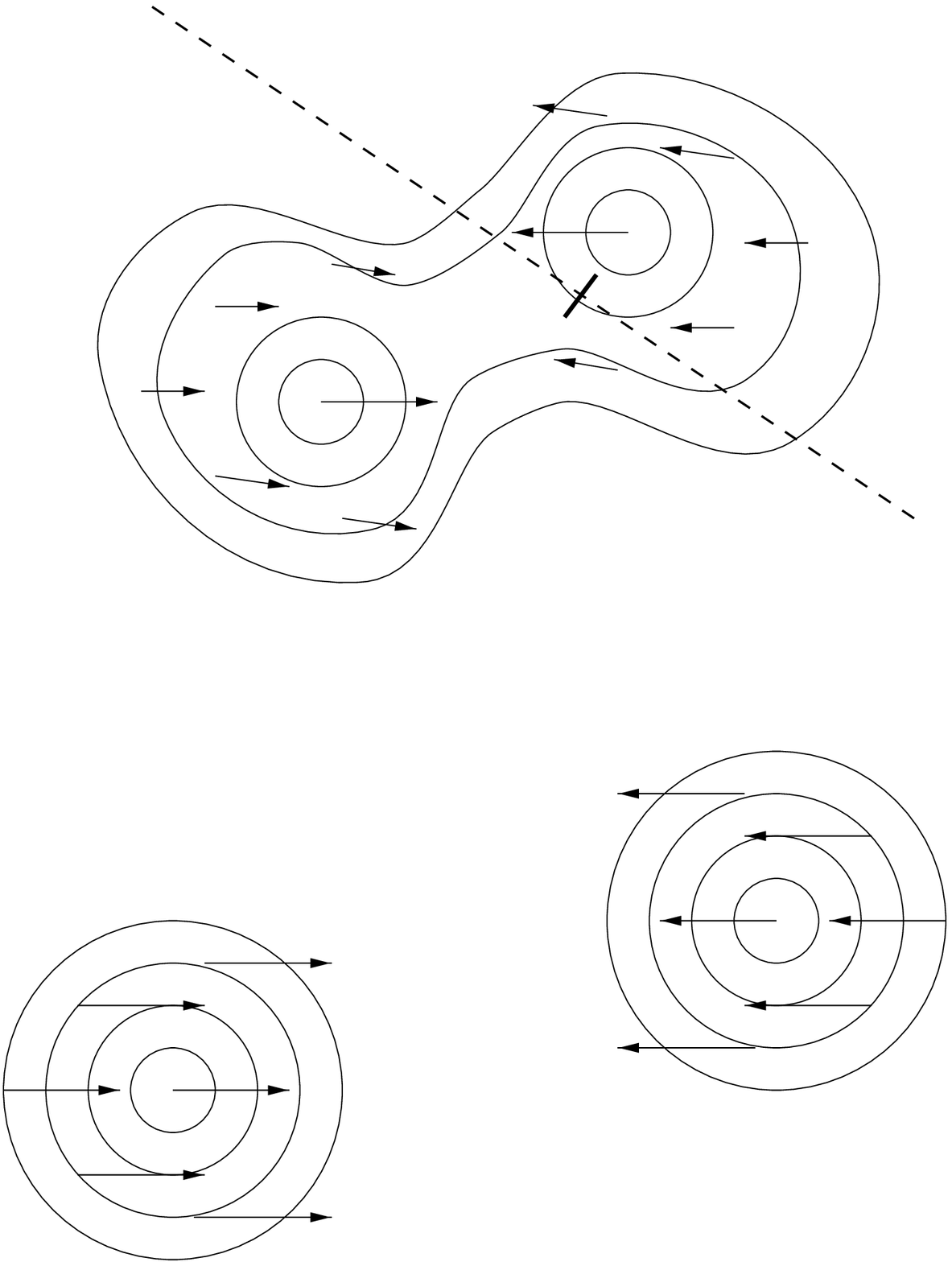}{12.0cm}{-90.0}{0.0cm}{2.cm}{.5cm}{5}
{A schematic view how leading edges arise 
from merging protogalactic clumps. The dashed line represents a random
line-of-sight with the expected position of the strongest absorption
feature marked.}
\end{figure} 

\subsection{How do asymmetric and leading edge profiles arise?} 

The main motivation for interpreting observed absorption profiles as 
a signature of rotation are their  leading edges,{\it i.e.,} 
the strongest absorption  feature often occurs at one of the edges of 
the profile. As demonstrated 
by Prochaska \& Wolfe (1997) such profiles occur naturally in  
a thick disk model  with an exponential
density  and an isothermal velocity profile.  
Fig.~5 shows such an example. There is, however, an equally simple and
plausible explanation  for such leading edges in a scenario of merging
protogalactic clumps. In the case of two  merging clumps
the strongest  absorption feature will generally be caused by the
high density central region of  the clump  which is closest to the 
LOS.  The  dense regions of merging PGC will move faster than 
their surroundings because of  the smaller deceleration by the 
(density-dependent) ram pressure. The strongest absorption feature 
then occurs naturally at the edge of the absorption profile. 
Smaller features are produced by density fluctuations in stripped 
material behind or by shocked material in front of the dense region.
This situation is illustrated 
schematically in Fig.~9.   Except  in the rare case where the LOS 
passes both dense regions symmetrically this is an intrinsically 
asymmetric configuration in velocity space.  This is 
the principal reason for the asymmetric LIS absorption profiles
in our models. 
We  caution, however, against over-interpretation of 
leading edge profiles. One should keep in  
mind that in the case of three randomly ordered components
of varying strength the probability that the strongest component is at
one edge is 2/3 and in the case of four components it is still 1/2.

\section{Absolute velocity widths and cosmological models.} 

In section 3 we have shown that the  width distribution
of the simulated LIS profiles relative to its median value
is consistent with that observed, 
and in section 4.1 we demonstrated that the width of the
profiles is correlated with the  
the  virial velocity of the associated DM halo.  
Fig.~10a shows the complete velocity width  distribution of the 
640 DLAS in our sample.
The median value is about 60 percent of  the  virial velocity of 
the associated DM halo, 
\begin{equation}
{\rm median}(\Delta v) \approx 0.6 \times  v_{\rm vir}.
\end{equation}
This value depends somewhat on the assumed  density threshold 
for self-shielding, the assumed metallicity, the selection 
criterion of the PGC and the redshift of the sample. Varying these
parameter we found the ratio of velocity width to virial velocity 
to vary between 0.5 and 0.75. 
One should note here that  the virial velocity 
which we infer from a given velocity width is a factor 1.5 to 2.5  
times smaller than in the rotating disk model.

\begin{figure}[t]
\plotmgh{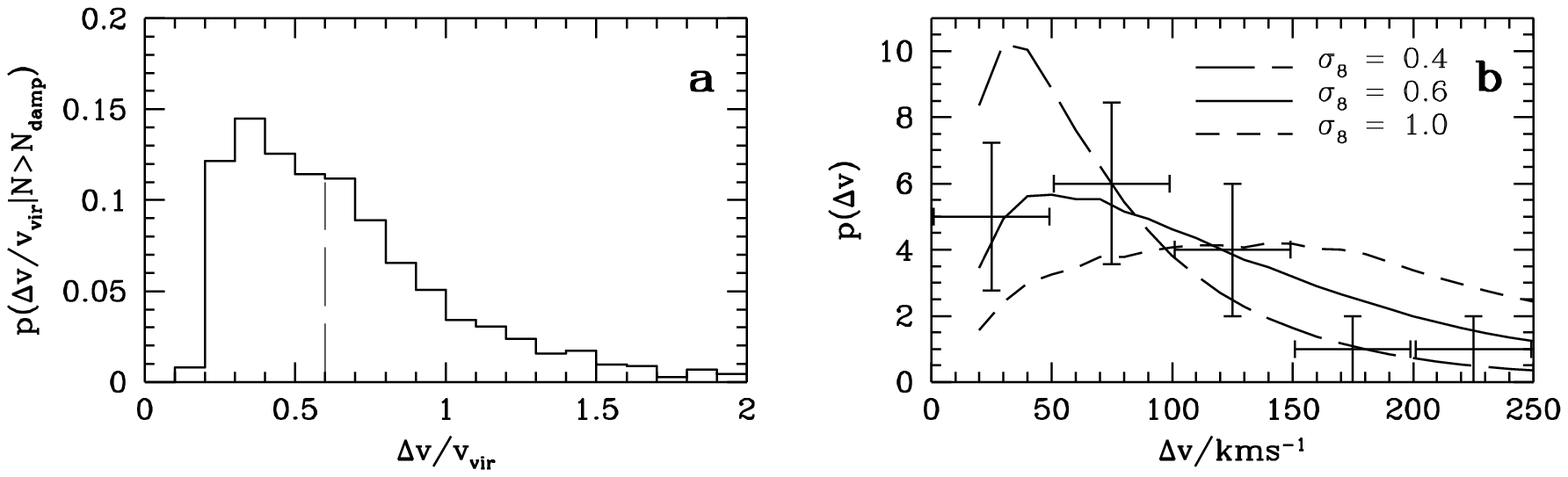}{17.0cm}{0.0}{-2.0cm}{-0.75cm}{-12.5cm}{6}
{\hfill  \break
a) Probability distribution of the velocity width scatter relative to
the virial velocity of the associated DM halo of the PGC giving rise
to the damped absorption system for 640 random LOS. The dashed line 
indicates the median value of 0.6.\hfill\break   
b) Probability distribution of absolute velocity widths at $z=2.1$ 
for CDM models with varying $\sigma_{8}$ as indicated on the plot. 
Crosses are the observed values from Prochaska \& Wolfe (1997).}
\end{figure}

It is difficult to assure that the sample of PGCs picked from our
numerical simulations is fully representative of the simulated
cosmological model. Furthermore it is very CPU time-consuming to simulate a
large number of different cosmogonies. The exact velocity width 
distribution will depend on the  
distribution of virial velocities in a chosen cosmological 
model weighted by the cross-section for damped absorption,
\begin{equation}
p(\Delta v,N_{\rm HI}>N_{\rm damp}, v_{\rm vir}) = 
p(\Delta v | N_{\rm HI}> N_{\rm damp},v_{vir})\times
p(N_{\rm HI}> N_{\rm damp} | v_{vir})\times
p(v_{vir}).
\end{equation}
We take the following
approach to calculate the distribution of absolute velocity widths. The
third factor in equation (8) the relative number of halos with 
different virial velocities is calculated using  the Press-Schechter 
formalism (Press \& Schechter 1974).
The cross section for damped absorption is assumed to scale linearly 
with mass,  $p(N_{\rm HI}> N_{\rm damp}|v_{vir})\propto 
M \propto v_{\rm vir}^{3}$. This is the simplest possible scaling 
suggested by the constant column density threshold 
defining a DLAS. A good estimate for the first factor is obtainable from our
numerical simulations. We found that  
$p(\Delta v | N_{\rm HI}>N_{\rm damp}, v_{\rm vir})$ 
depends mainly on the ratio of $\Delta v/v_{\rm vir}$
and only  weakly on the virial velocity of the dark matter halo
itself. We therefore used the  velocity width  distribution  of all  
640 DLAS shown in Fig.~10a.  The distribution of
absolute velocity widths is obtained by integrating over virial velocity,  
\begin{equation}
p(\Delta v,N_{\rm HI}>N_{\rm damp}) =
\int_{v_{\rm min}} ^{\infty} 
{p(\Delta v , N_{\rm HI}> N_{\rm damp},v_{vir}) \dd v_{\rm vir}}.
\end{equation}
The result is shown is shown in Fig.~10b for a standard CDM model 
at $z= 2.1$ with three different values  of $\sigma_{8}$
and a minimum virial velocity of 30 $\kms$.  
Even the largest observed velocity width, $\sim 200 \kms$, can 
be accommodated in a CDM model  with $\sigma_{8}$ as low as 0.5 to
0.6.  Most  of the currently
favored variants of hierarchical galaxy formation should therefore
have no serious problems in account for the observed 
velocity width distribution.   
Constraints on different   hierarchical  scenarios by the overall 
incidence rate for damped absorption have been discussed extensively
by other authors (see also section 6).

\section{Damped \op absorber --- Large disks or protogalactic clumps?}

Current hydrodynamical simulations, including those 
presented here, are undoubtedly unable to model all the details of the   
spatial distribution and  kinematics of the gas in the innermost 
regions of  collapsed dark matter halos. We believe, however, 
that our simulations  already catch many of the significant features
and as dicussed above they should underestimate  rather than 
overestimate  the amount of structure in the density 
and velocity field. The simulations therefore 
clearly cast serious doubt on the claim  that 
only objects as massive as present-day sprials can produce
the velocity widths of the  observed LIS profiles, and 
that rotation is the only  possible  interpretation for the shape
of the profiles.

Hierarchical structure formation models can explain many other
features of the cosmic matter distribution 
seen in absorption, e.g., the rate of incidence, column density 
distribution, Doppler parameters, ionization state, sizes, and the 
opacity of Lyman $\alpha$ forest clouds, and the abundance, kinematics,
temperatures, and ionization conditions
of heavy element absorbers (Cen et al.~1994; Hernquist et al.~1996; 
Petitjean, M\"ucket \& Kates 1995;  Miralda-Escud\'e et al.~1996;  
Zhang et al.~1997;  Croft et al.~1997; Rauch et al.~1997,
Bi \& Davidsen 1997; Hellsten et al.~1997; paper I\&II). 
The fact that the same models can account for the essential features of DLAS (Katz et al.~1996;
Gardner et al.~1997a/b; paper I\&II) makes absorption by protogalactic
clumps an even more  attractive explanation for DLAS.
However, some of the observed properties of absorption systems are 
not unique to hierarchical models and  we do not consider the rapidly 
rotating, large disk hypothesis to be ruled out by our results. Here we
briefly comment on other arguments  which have been put forward in  
discussing  the nature of DLAS, mostly in favor of DLAS as large, protogalactic
disks:

\noindent (1) The high column densities:

Large \hi\ column densities of DLAS are indeed reminiscent of present-day
disks (e.g. Wolfe 1988). However, simple analytical estimates and the 
various simulations quoted above show that these column densities 
can equally well be  produced in gas-rich  protogalactic clumps
with masses expected in typical hierarchical structure formation models.  

\noindent (2) The large impact parameter:

Large separations between  the absorber 
and detected  emission attributed to   associated starlight
are taken as indicative of extended, massive objects. However, 
very few examples
are currently known (M\o ller \& Warren 1995; 
Warren \& M\o ller 1995, Djorgovski 1996). It is not yet clear whether the
inferred sizes actually contradict the predictions by 
hierarchical structure formation scenarios (Mo, Mao \& White 1997). 
This will depend crucially on the detailed gas distribution in the 
outskirts of protogalactic clumps. Moreover,  there are several 
uncertainties.
We do not know yet how emission and absorption properties are
related.  As demonstrated  in section 2 and papers I\&II the region 
responsible for DLAS often 
contains several protogalactic clumps within a few tens of kpc.
Also, the few DLAS identified in emission may only 
be the tip of the iceberg, at the upper end of the mass and size  
distribution. 

\noindent (3) The continuity of $\Omega_b$:

$\Omega_b$ in the \hi\ phase of high-redshift DLAS is roughly similar
to the baryon content in the stellar component of present-day spirals
(Wolfe 1986;  Lanzetta et al.~1995; Storrie-Lombardi 1996).  This has
been interpreted as continuity in baryon content between DLAS and
present-day galaxies. If correct this continuity  may mean that the gas
constituting the stars observed today has already cooled and settled
into collapsed objects at these redshifts. 
The coincidence contains, however, no information on the size 
distribution of the collapsed objects.  Hierarchical models have been
shown to reproduce the observed $\Omega_b$ in DLAS and its evolution 
with redshift (Kauffmann \& Charlot 1994; Ma \& Bertschinger 1994; Mo \&
Miralda-Escud\'e 1994;  Klypin et al. 1995; Gardner et al.~1997a/b; Ma
et al.~1997).

\noindent (4) The high rate of incidence:

The rate of incidence only depends on the product of space density of 
the absorbers times their cross section for damped absorption. It
gives  no constraints on the size of individual absorbers. 
Hierarchical models have also  been shown 
to reproduce the observed rate of incidence of DLAS and 
its evolution with redshift (Kauffmann 1996; Baugh et al.~1997).

\noindent (5) The alignment of the edge of the absorption profile with the
redshift of observed emission:

For two of the DLAS where emission has been observed it is  possible 
to  determine   the relative position of  emission and 
absorption in redshift space (Lu, Sargent \& Barlow 1997). 
In the first case the emission falls on one edge of the LIS absorption profile 
in absorption while the strongest absorption feature lies
on the opposite edge. In the second case the emission redshift lies
roughly at the centre of the absorption profile. 
If the LIS absorption profiles showing the 
leading-edge signature were solely due to rotation,  
if, in addition, the center of emission coincided with the center of the
disk and if finally there were no optical depth effects, then the 
emission redshift should indeed occur  preferentially  at the edge of the 
LIS absorption profile opposite to the edge coinciding with the strongest 
absorption feature. The situation is, however, similar if the leading 
edges are due to  merging/collision of PGCs. The emission of a
stellar continuum would most likely originate in one of the central
regions of the merging clumps and should therefore also coincide with one 
of the two edges of the absorption profile. However, we see no reason
why that should happen preferentially opposite to the edge with the
strongest absorption feature. Thus, a large number of cases like that
reported by Lu et al.~would indeed argue for rotation 
dominating the dynamics.

For completeness, we mention a counterargument against massive disks
that has received some attention in the past,  the issue of the
metallicities in DLAS.  The metal abundances ([Fe/H]) in DLAS at high z
are much lower than expected for local spiral disks (Pettini et al.
1994; Lanzetta, Wolfe, and Turnshek (1995); Lu et al.~(1996); Prochaska
\& Wolfe 1996), which has led to suggestions that DLAS show abundance
patterns of dwarf galaxies or galactic halos. 
In the CDM picture, this inference is correct.

\section{Conclusions}

We have used hydrodynamical simulations of galaxy formation 
in a cosmological context to study the line profiles of low ionization species 
associated with damped \op absorption systems. Observed velocity
widths and asymmetries
of the line profiles of the low ionization species
are well  reproduced by a mixture of rotation, random motions, infall and
merging of protogalactic clumps.  
The asymmetries are mainly caused by random sampling of irregular density
and velocity fields of individual halos and by intrinsically  asymmetric
configurations arising when two or more clumps collide.  
We show why  leading-edge  asymmetries occur  naturally in the latter
case; the dense central regions of the clumps  move faster than  
surrounding less dense material.

We have further shown that the presence of non-circular motions reduces 
the depth of the potential well necessary to produce a given velocity
width   compared to a model where the absorption  is solely due to
rotation.  The reduction is typically a factor about 2.
The observed velocity width can therefore be explained by gas moving into and
within (forming) dark matter halos 
with typical virial velocities of about $100\kms$. Velocity width and 
virial velocity are linearly correlated but  the scatter is large; there
are outliers with large velocity width due to an occasional alignment 
of clumps with well separated dark matter halos.   

Our final conclusion is that asymmetric profiles of the kind observed are 
not necessarily  the signature of rotation and that there is no problem 
of accommodating  the observed velocity widths within standard hierarchical
cosmogonies.

\noindent

\section{Acknowledgments}

We thank Simon White for a careful reading of the manuscript, and
Limin Lu, Hojun Mo, Jason Prochaska, and Art Wolfe for very helpful 
discussions. MR is grateful to NASA for support through
grant HF-01075.01-94A from the Space Telescope Science Institute,
which is operated by the Association of Universities for Research in
Astronomy, Inc., under NASA contract NAS5-26555.
Support by NATO grant CRG 950752 and the ``Sonderforschungsbereich 375-95
f\"ur Astro-Teilchenphysik der Deutschen  Forschungsgemeinschaft''
is also gratefully acknowledged.

\pagebreak

\pagebreak

\end{document}